\documentclass[%
 reprint,
 amsmath,amssymb,
 aip,
 apl,
]{revtex4-2}

\usepackage{graphicx}
\usepackage{color}
\usepackage{slashbox}
\usepackage[colorlinks=true,linkcolor=blue]{hyperref}
\usepackage{dcolumn}
\usepackage{bm}

\begin{document}

\title{The effect of confinement on thermal fluctuations in nanomagnets}
\author{Samuel D. Sl\"oetjes}
 \email[Author to whom correspondence should be addressed: ]{samuel.sloetjes@physics.uu.se}
 
\author{Bj\"orgvin Hj\"orvarsson}

\author{Vassilios Kapaklis}%
 \email[Author to whom correspondence should be addressed: ]{vassilios.kapaklis@physics.uu.se}

\affiliation{Department of Physics and Astronomy, Uppsala University, Box 516, SE-75120 Uppsala, Sweden}

\date{\today}

\begin{abstract}
We study the magnetization dynamics in nanomagnets 
excited by stochastic magnetic fields to mimic temperature in a micromagnetic framework. 
The effect of confinement arising from the finite size of the structures is investigated, and we visualise the spatial extension of the internal magnon modes.  
Furthermore, we determine the temperature dependence of the magnon modes, and focus specifically on the low frequency edge modes, which are found to display fluctuations associated with switching between C- and S-states, thus posing an energy barrier. We classify this fluctuating behaviour in three different regimes and calculate the associated energy barriers using the Arrhenius law.

\end{abstract}

\maketitle

Mesoscopic spin systems can be used as a playground for investigations of magnetic ordering and dynamics \cite{Nisoli2013,Heyderman:2013gb,Nisoli_NatPhys_Perspective,Rougemaille_Colloquium_2019}.  A range of mesoscale magnetic structures have been fabricated using nanolithography, spanning from analogues to the 1D and 2D Ising model systems\cite{Arnalds_2DIsing, Ostman_Ising_2018}, to extensive two-dimensional frustrated artificial spin ice (ASI) structures \cite{Wang2006, Perrin_Nature_2016,Ostman_natphys_2018}. The elements are often treated as point-like magnetic dipoles and more recently as artificial {\it magnetic-atoms}. These analogies only hold to a certain point when describing thermal fluctuations and transitions in mesoscopic systems. Furthermore, it has become evident that the analogy to a point-like dipole can even be misleading, resulting in misinterpretations and quantitative discrepancies between observations and calculations \cite{ThermalFluctuationsInASI,Andersson2016,Morley2017,ASI_susceptibility_2020}. The reason for this originates in contributions from both static and dynamic textures in the magnetisation of the elements \cite{bayer2005spin,Bessarab2012,Bessarab2013,Gliga_PRB_2015}. Even though extensive work has been done  \cite{Gliga_PRL_2013,iacocca2016reconfigurable,iacocca2020tailoring,Gliga_PRB_2015}, little is known about the effect of the extension of the elements on the thermal excitation. 

\begin{figure}[t]
    \centering
    \includegraphics[width=\columnwidth]{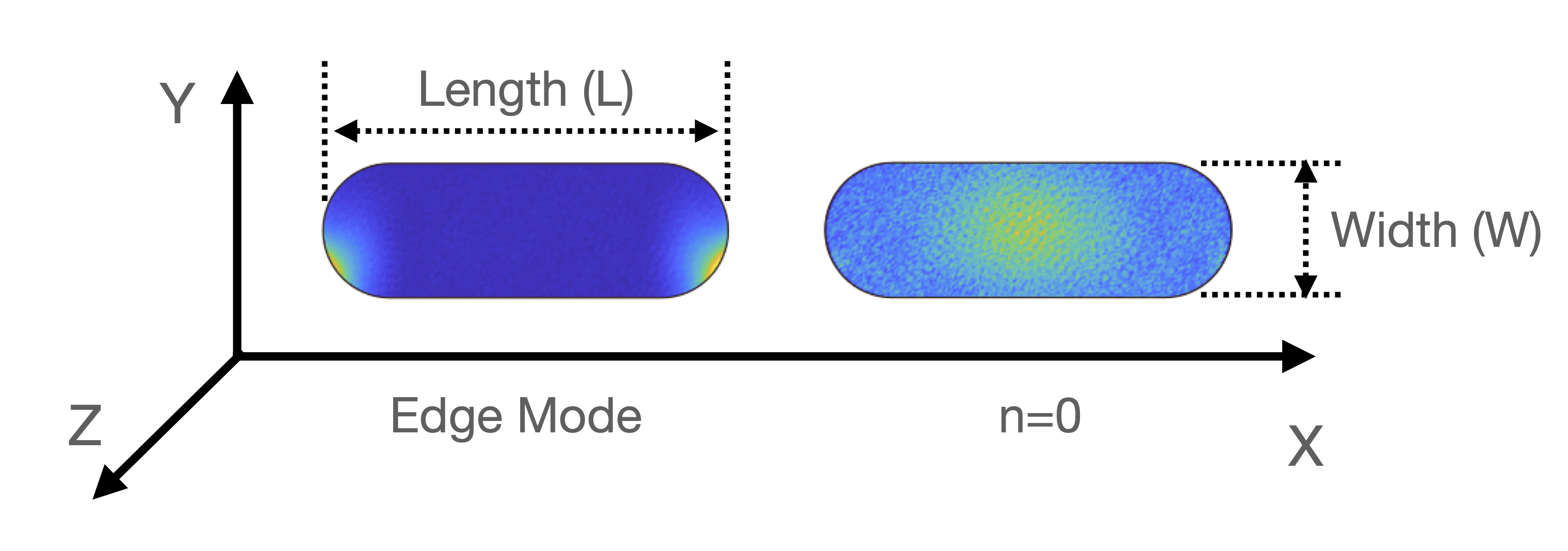}
    \caption{Schematic illustration of  mesospins studied and the spatial coordinate system. The mesospins have a stadium-like shape, common for Ising mesospins, with lengths $L$ = 450, 360 and 270 nm and aspect ratios of 90:30:1 in all cases. The color shading denotes the fluctuation amplitude for the perpendicular magnetization component, $m_z$, for two of the characteristic modes which are thermally excited (see Fig. \ref{fig:AllSpectra} for a complete list of modes).}
    \label{fig:Fig1}
\end{figure}

Here we investigate the influence of temperature on the inner magnetisation of the Ising like mesospins. 
The model system we use for these investigations consists of elongated, stadium-shaped nanomagnets, as illustrated in Fig. \ref{fig:Fig1}, with an aspect ratio of length ($L$):width ($W$):thickness ($t$) $= 90:30:1$. Henceforth, we will refer to these magnetic elements as {\it mesospins}. We use the micromagnetic simulation package MuMax$^3$ for all the calculations\cite{mumax3}. Effects such as exchange, crystalline anisotropy, and demagnetization are taken into account by means of an effective field. The mesospins are assumed to have the same magnetic properties as Permalloy (Py), with a saturation magnetization of $M_s=10^6$~A/m, and an exchange stiffness of $A_{ex}=10^{-11}$~J/m. The Gilbert damping constant is set to $\alpha=0.001$. The structure is divided up in cells, the size of which are given by $l_x \times l_y \times l_z =$ 2.5~nm~$\times$~2.5~nm~$\times$~$t$~nm. The in-plane component of the cell size is smaller than the exchange length, given by $l_{ex} = \sqrt{2A_{ex}/\mu_0M_S^2}=4.0$~nm, ensuring reliable simulation results \cite{mumax3}. A typical problem that appears when simulating magnetization dynamics in a micromagnetic framework, is the appearance of a Van Hove-singularity, resulting of the discretization of the magnetic continuum, and the corresponding cut-off for spin waves with a wavelength smaller than twice the cell size\cite{berkov2005stochastic}. However, with the current cell size, we find that this singularity is moved far beyond 100 GHz, i.e. much larger than the frequency region of interest ($f$~$<$~30~GHz).

The temperature is simulated by a time varying thermal field $\mathbf{H}_i^{therm}(t)$, with the following properties:
\begin{align*}
    \label{eq:therm}
    \langle\mathbf{H}_i^{therm}(t)\rangle &= 0\\
    \langle\mathbf{H}_i^{therm}(0)\mathbf{H}_j^{therm}(t)\rangle &= 2D\delta(t)\delta_{ij}
\end{align*}
where the $i$ and $j$ denote cell indices, $D$ is the power of the fluctuations and $\delta$ Kronecker's delta function. The first equation implies that the thermal field vanishes upon averaging. The second equation defines delta correlations both in space and time. The delta correlation in time is justified by considering that the correlation time of the thermal fluctuations is a few picoseconds, i.e. of the order of the inverse Debye frequency, as thermal fluctuations of the spin degrees of freedom originate from interaction with phonons. This timescale is much smaller than that typical of magnetization dynamics. The delta correlation is space is justified by considering that the correlation length of thermal fluctuations is typically a few unit cells, i.e. much smaller than the micromagnetic cell size\cite{bayer2005spin,berkov2005stochastic}. As such, the thermal field in a micromagnetic framework is effectively random in space and time, which is numerically realized through a random vector $\boldsymbol{\eta}$, the size of which varies with a Gaussian distribution around unity, and whose direction is randomized for every timestep and cell. The fluctuation power $D$ can be found from the fluctuation-dissipation theorem\cite{brown1963thermal}, thus the expression for the thermal field becomes:
\begin{equation}\label{eq:therm}
    \mu_0\mathbf{H}^{therm} = \boldsymbol{\eta}\sqrt{\frac{2\alpha k_BT}{M_s\gamma V \Delta t}}
\end{equation}
\noindent
here, $\alpha$ is the Gilbert damping constant, $k_B$ is the Boltzmann constant, $T$ is the temperature, $M_s$ is the saturation magnetization, $\gamma$ is the gyromagnetic ratio, $V$ is the volume of the cell, and $\Delta t$ is the time step. A sixth order Runge-Kutta-Fehlberg solver is used in MuMax$^3$ to calculate the thermal fluctuations using adaptive time steps \cite{Leliaert_thermal_mumax3}. The spatial and temporal randomness of the field ensures excitation of all eigenmodes in the structures, as opposed to methods where more homogeneous magnetic fields are used for the excitations \cite{mcmichael2005magnetic}.

The time window of the simulations is typically 25~ns, within which the magnetization vector $\mathbf{m}(x,y,t)$ is recorded every 5 ps, resulting in a frequency resolution of 0.04 GHz and a range of 0-100 GHz. In order to obtain reliable spectra, each simulation is run four times with different thermal seeds, after which the resulting spectra are averaged. The spatial dependence of the magnon amplitudes can be found by taking the Fourier transform of fluctuating components via $m_{y,z}(x,y,f)=\mathcal{F}\{m_{y,z}(x,y,t)\}$ \cite{ciuciulkaite2019collective, sloetjes2019effects}. Furthermore, the spatial dependence can be averaged out in order to obtain the spectrum, via $\langle m_{y,z}(x,y,f) \rangle_{x,y} = m_{y,z}(f)$. The magnon spectral density $n$, can be extracted from $m_{y,z}(f)$, by using the following relation \cite{gurevich1996magnetization}: $n(f) = |m_{y}(f)|^2 + |m_{z}(f)|^2$. 



\begin{figure}[t]
    \centering
    \includegraphics[width=\columnwidth]{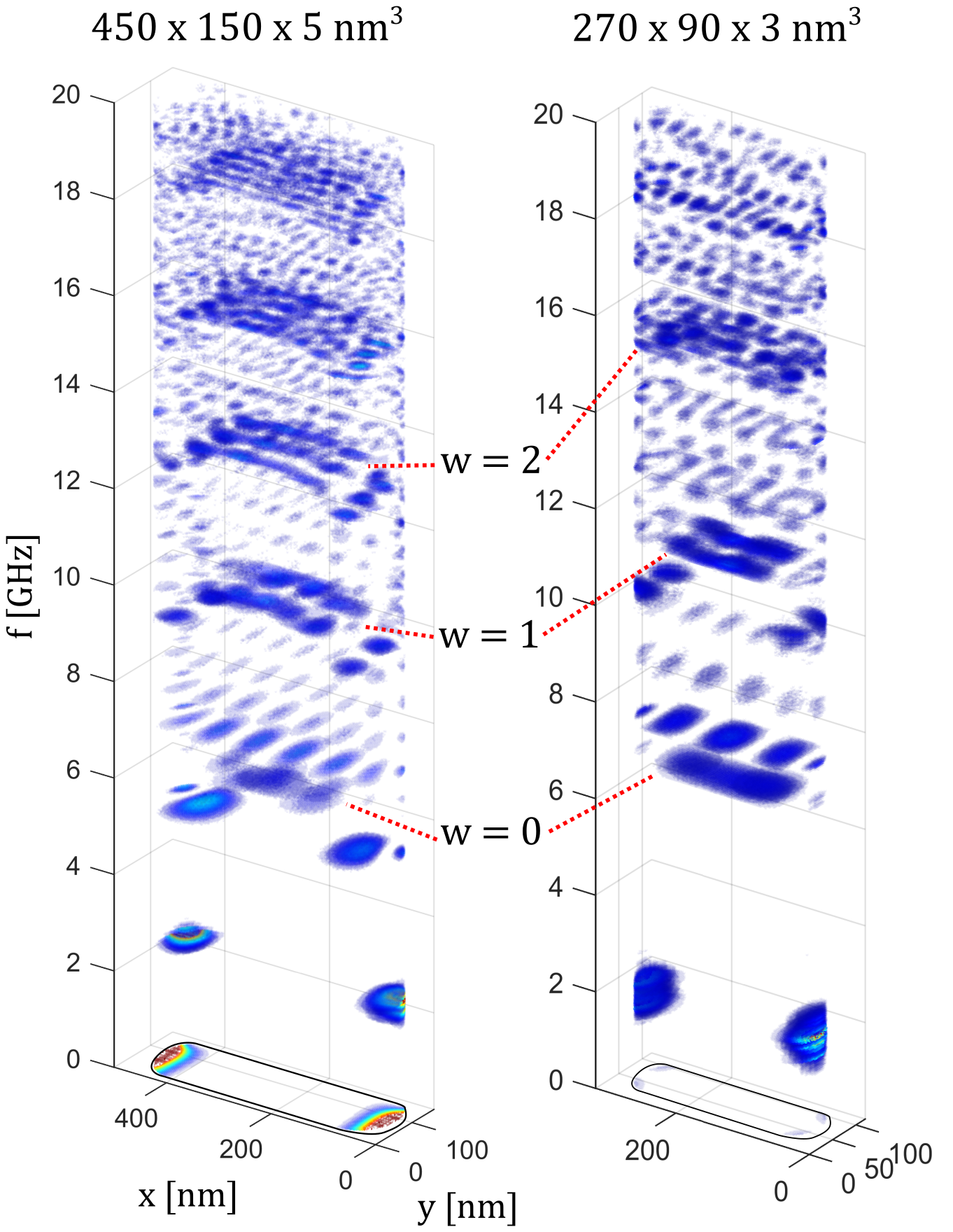}
    \caption{Spatial dependence of the thermal magnon intensity for two different mesospin sizes in the frequency range 0-20~GHz.}
    \label{fig:AllSpectra}
\end{figure}

In the thermodynamic limit, the magnon spectrum is continuous for isotropic ferromagnets. When the size is finite, a gap will be obtained at $|\mathbf{k}|=0$. Fig. \ref{fig:AllSpectra} shows the full spectrum of magnons per unit area for two different mesospin sizes, taken at $T$ = 100 K. Standing magnon modes emerge in the longitudinal and transversal directions, the order of which we indicate with the integers $v$ and $w$, respectively. The uniform $(v,w)$~=~$(0,0)$ mode shows up at $f$~=~6.3~GHz for the mesospin with $L$~=~450~nm, and splits into higher order longitudinal modes as the frequency is increased. One exception to this is that the (1,0)-mode has a lower frequency than the uniform mode, whereas all the other modes with $v>$~1 have a frequency higher than the Kittel mode. This is a consequence of the dynamic dipolar interaction in the case that $\mathbf{k} \parallel \mathbf{m}$. In this configuration, the dispersion relation, $f(\mathbf{k})$, has a minimum for $\mathbf{k}\neq0$, i.e. a magnon with a finite wavelength has the minimum frequency. The frequency gaps between the transverse magnon modes are much larger than the gaps for the longitudinal modes, as a result of difference in extension. In addition to the modes in the interior of the elements, we observe edge modes, the lowest order of which are seen at $f$ = 2.5 and 5.5 GHz. The $L$~=~270 nm mesospin shows only a single edge mode, centered around 1.8 GHz. 

\begin{figure}
    \centering
    \includegraphics[width=\columnwidth]{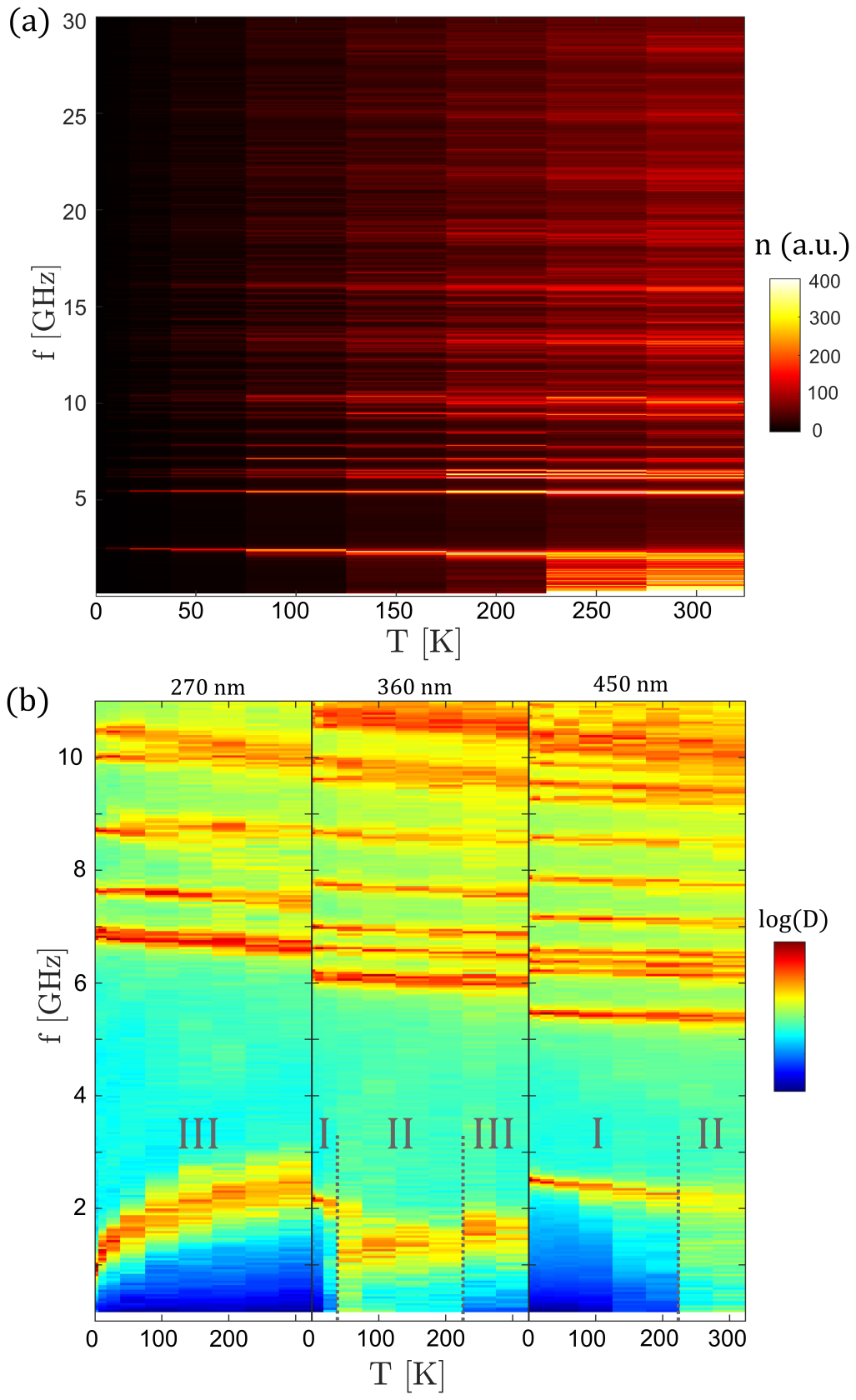}
    \caption{(a) Magnon spectral density for the $L$ = 450 nm mesospin as a function of temperature and frequency. (b) The magnon occupation numbers as a function of temperature and frequency for three different mesospin sizes. The gray lines serve as guides to the eye for the transitions between different regimes, labeled by roman numerals.}
    \label{fig:Tmap}
\end{figure}

An increase in temperature leads to an increase of occupied magnon modes, as shown in Fig. \ref{fig:Tmap}a, and at low temperatures, only the lowest lying states are occupied. Between 250 and 300 K, we observe an increased occupation of states in the gap region at $f<3$~GHz. To get a better picture of the change of available states with temperature, we investigated the magnon occupation numbers (MON). We obtain this quantity using $n(E,T) = D(E,T)F(E,T)$, where $D(E,T)$ are the magnon occupation numbers, and $F(E,T)$ is the thermal distribution function. Magnons are bosons, following Bose-Einstein statistics. However, since each cell in the micromagnetic simulation is a coarse grained average over a large ensemble of quantum mechanical spins, a classical description of the cells should be sufficient. Therefore, we use the Rayleigh-Jeans distribution, which scales as $F(E,T) \propto T/E$ \cite{ruckriegel2015rayleigh}. The MON are calculated using $D(E,T) = n(E,T)/F(E,T)$ and the results are plotted in Fig. \ref{fig:Tmap}b. For frequencies $f>5$ GHz, we observe a slight decrease in the resonance frequencies with increasing temperature, which likely results from a decreased effective field due to a lower overall magnetization, a mechanism which is captured by Bloch's law. Additionally, mode hybridization occurs for two modes located around 10 GHz. Using amplitude maps, we find that the mode with the lower frequency is a center mode and the higher frequency mode is an edge mode.

The most striking difference in the temperature dependence of $D(E,T)$ can be seen in the low frequency region, i.e. $f<3$ GHz, which is populated exclusively by edge modes. We can discern three different thermal regimes for the behaviour of these edge modes, the numbers of which are indicated in Fig. \ref{fig:Tmap}b. Below 200 K, the mesospin with $L$~=~450~nm features a low frequency mode at 2.4 GHz which decreases slightly in frequency as the temperature is increased. At $T$ $>$ 200 K, we observe the emergence of additional states spanning the range 0 to 2 GHz, which implies a transition between two different regimes. A third regime can be identified, as the $L$~=~270~nm mesospin features a mode at a similar position that increases significantly with frequency as the temperature is increased, with no available states below these frequencies. The mode moves from 1 GHz at low temperatures to 2.4 GHz at 300 K, and the peak position of this mode scales as $f\propto T^{\frac{1}{4}}$. 
We observe for this particular mode that the ellipticity decreases with increasing temperature (see Suppl. Material). For the mesospin with $L$~=~360~nm, we find behaviour indicative of a transition between these three regimes, the first transition happening at $T$~=~10~K, and the second happening at $T$~=~200~K. 

\begin{figure}[t]
    \centering
    \includegraphics[width=\columnwidth]{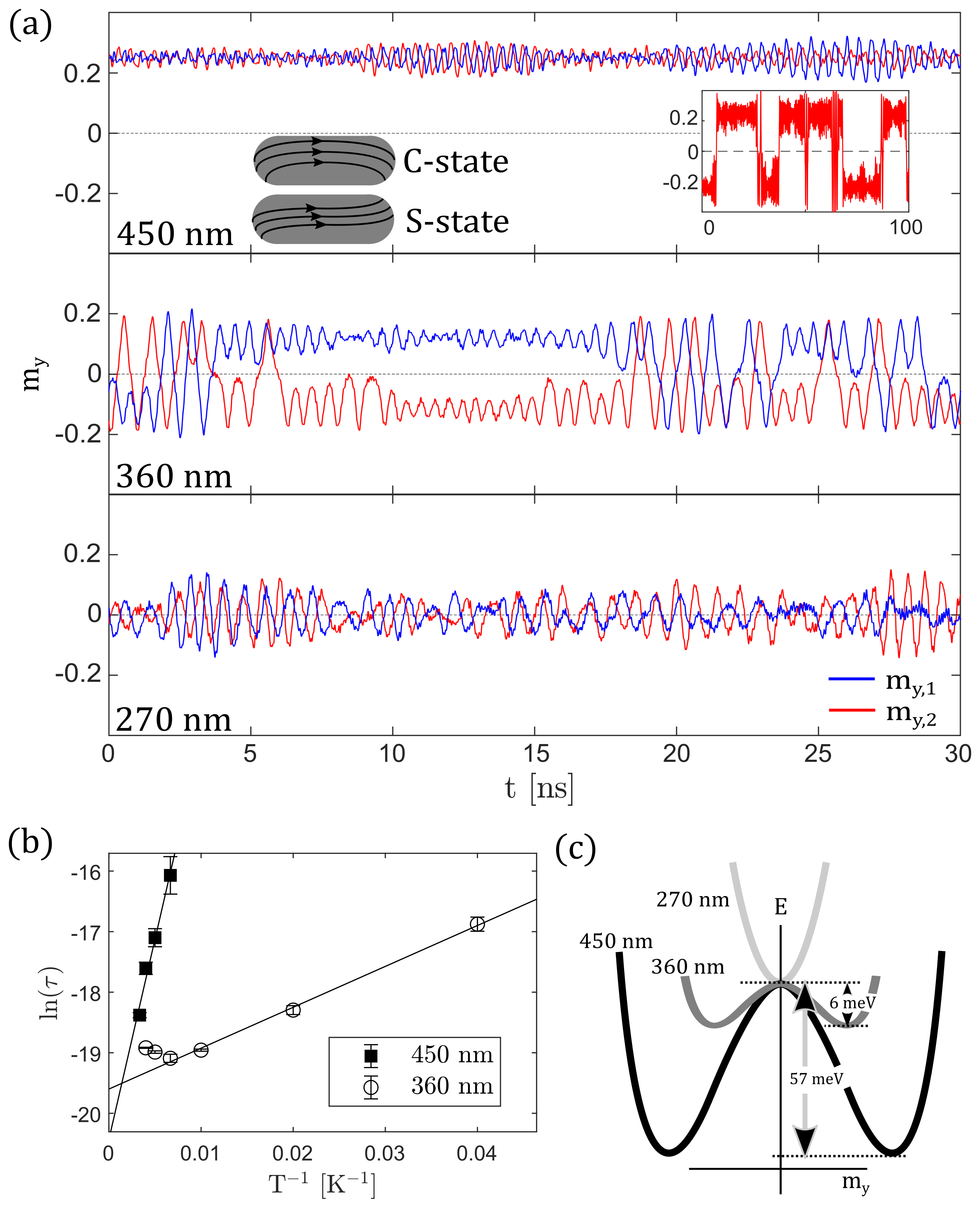}
    \caption{(a) Fluctuation of the $m_y$-component of the two edges of the mesospin at $T$ = 50 K, denoted by $m_{y,1}$ and $m_{y,2}$, for different sizes. The inset shows the switching of one edge of the $L$ = 450 nm mesospin at a raised temperature of 250~K. (b) Arrhenius plot of the average relaxation time of the edge fluctuations, at different temperatures, for two different sizes. (c) A schematic pointing out the energy landscape for the different mesospin sizes, and the magnitude of the energy barriers extracted from the linear fits in (b).}
    \label{fig:flucs}
\end{figure}

In order to uncover the origin of these transitions, we inspect the averaged transverse $m_y$ components at the edges of the $L$~=~270, 360 and 450 nm mesospins, as shown in the upper, middle and lower panel of Fig. \ref{fig:flucs}a, for a temperature of $T=50$ K. At this temperature, the mesospin with $L$ = 450 nm is in Regime I, and can be seen to oscillate around a non-zero value of $m_y$. We interpret this result as the mesospin being locked into either a C- or an S-state (see inset of top panel in Fig. \ref{fig:flucs}a), where it precesses. The mesospin with $L$ = 360 nm is in Regime II, which is characterized by irregular switching of the edge magnetization in the transverse direction, which occurs at longer timescales than the precessional motion in the locked C- or S-state (see Fig. \ref{fig:flucs}a, middle panel, and Suppl. Material). This slow switching process explains the increased intensities at lower frequencies in the magnon occupation numbers (see Fig. \ref{fig:Tmap}b, right panel). These two distinct regimes imply the presence of an energy barrier, the amount of transitions over which is determined by the temperature. The $L$~=~270~nm mesospin is in Regime III over the whole temperature interval, and oscillates constantly around $m_y=0$, as illustrated in the lower panel in Fig. \ref{fig:flucs}. This behaviour is consistent with the absence of an energy barrier between C- and S-states for this mesospin size, meaning that the $L$~=~270~nm mesospin does not have an S- or C-state configuration as a groundstate. It is thus a balance between demagnetization energy and exchange energy, which determines whether an energy barrier is formed, a line of reasoning that is similar to flux closure/single domain magnetization transitions in mesoscopic structures of low aspect ratio \cite{ding_magnetic_2005}.

We can estimate the height of the barriers in the $L$~=~360~nm and $L$~=~450~nm mesospin using Arrhenius law ($\tau = \tau_0 e^{\frac{\Delta E}{k_BT}}$), as illustrated in Fig. \ref{fig:flucs}b. Here, $\tau$ is the inverse switching rate, given by the average time spent in either configuration having a positive or negative $m_y$, $\tau_0$ is the inverse attempt frequency, and $\Delta E$ is the height of the energy barrier. The variable $\tau$ can be found by dividing the total simulation time by the amount of switches. One should in this case be careful not to take into account `false' switching events, i.e. the edge magnetization must spend sufficient time in a metastable state\cite{Semenova_PRB_2020}. We demand that the time it takes to equilibrate should be larger than $\tau_{eq}~=~1$~ns, 
and disregard switching events that occur within a shorter interval after an initial switching event. 
A long simulation of 1 $\mu s$ was performed in order to obtain sufficient statistics on the switching. The uncertainty is determined from the deviation of switching rates between the two different edges. By fitting the Arrhenius law, we find a significant difference in the activation energy: 6~meV and 57~meV for the $L$~=~360~nm and $L$~=~450~nm mesospins, respectively. The inverse attempt frequencies are  $\tau_0$~=~3.08$\times$10$^{-9}$ s ($L$~=~360~nm) and $\tau_0$~=~1.35$\times$10$^{-9}$ s ($L$~=~450~nm). The energy landscapes, and the corresponding values for the energy barriers are illustrated in Fig. \ref{fig:flucs}c. The data points for the two highest temperatures belonging to the 360~nm nanomagnet are seen to deviate from the otherwise linear relation. This is due to too many false switching events, thereby masking true switching events, therefore we assign no weight to these data points in the fitting procedure. This shortcoming calls for more sophisticated methods to more accurately determine the dynamics of the switching. 

Additionally, we evaluated the correlation of the magnetic state of the edges, by calculating the Pearson correlation coefficient numerically, $\rho (m_{y,1},m_{y,2})$ as described in the Suppl. Material. All the tested temperatures and mesospin sizes show a weak anticorrelation within the range $-8$~\%~$<$~$\rho$~$<$~0~\%, except for the constant switching of the $L$~=~360~nm mesospin at 250~K ($\rho$~$=$~0.8~\%), and the $L$~=~450~nm mesospin at 50~K ($\rho$~$=$~-16~\%). The weak anticorrelation likely originates from the weak stray field interaction between the $m_y$ components, which favours oppositely aligned magnetization in the lateral direction. 



The observed fluctuations might play a strong role in the spectral response and symmetry breaking in vertices of ASI arrays, with temperature, as presented here, being a further tuning parameter \cite{Gliga_PRB_2015}. The theoretical and simulation approach described for addressing thermal excitations in mesoscopic magnetic systems, is potentially useful for resolving emergent collective behavior. The latter is particularly important for solving issues related to the ordering and thermal excitations of coupled mesospins \cite{ASI_susceptibility_2020, Mellado_arXiv_2020, leo2021chiral, Hofhuis_DM_ASI_2020}. This knowledge may even find its application in logic and computational applications \cite{Lendinez_review_2020}, such as design of neuromorphic-like architectures based on ASIs and their magnonic properties \cite{Gypens_2018}. 
\begin{acknowledgments}

The authors would like to acknowledge financial support from the Swedish Research Council (Project No. 2019-03581), the Swedish Foundation for International Cooperation in Research and Higher Education (Project No. KO2016-6889) and from the Carl Tryggers Foundation (Project No. 19:175).

\end{acknowledgments}

\section*{DATA AVAILABILITY}

The data that support the findings of this study are available from the corresponding author
upon reasonable request.


\begin{thebibliography}{37}
\makeatletter
\providecommand \@ifxundefined [1]{%
 \@ifx{#1\undefined}
}%
\providecommand \@ifnum [1]{%
 \ifnum #1\expandafter \@firstoftwo
 \else \expandafter \@secondoftwo
 \fi
}%
\providecommand \@ifx [1]{%
 \ifx #1\expandafter \@firstoftwo
 \else \expandafter \@secondoftwo
 \fi
}%
\providecommand \natexlab [1]{#1}%
\providecommand \enquote  [1]{``#1''}%
\providecommand \bibnamefont  [1]{#1}%
\providecommand \bibfnamefont [1]{#1}%
\providecommand \citenamefont [1]{#1}%
\providecommand \href@noop [0]{\@secondoftwo}%
\providecommand \href [0]{\begingroup \@sanitize@url \@href}%
\providecommand \@href[1]{\@@startlink{#1}\@@href}%
\providecommand \@@href[1]{\endgroup#1\@@endlink}%
\providecommand \@sanitize@url [0]{\catcode `\\12\catcode `\$12\catcode
  `\&12\catcode `\#12\catcode `\^12\catcode `\_12\catcode `\%12\relax}%
\providecommand \@@startlink[1]{}%
\providecommand \@@endlink[0]{}%
\providecommand \url  [0]{\begingroup\@sanitize@url \@url }%
\providecommand \@url [1]{\endgroup\@href {#1}{\urlprefix }}%
\providecommand \urlprefix  [0]{URL }%
\providecommand \Eprint [0]{\href }%
\providecommand \doibase [0]{https://doi.org/}%
\providecommand \selectlanguage [0]{\@gobble}%
\providecommand \bibinfo  [0]{\@secondoftwo}%
\providecommand \bibfield  [0]{\@secondoftwo}%
\providecommand \translation [1]{[#1]}%
\providecommand \BibitemOpen [0]{}%
\providecommand \bibitemStop [0]{}%
\providecommand \bibitemNoStop [0]{.\EOS\space}%
\providecommand \EOS [0]{\spacefactor3000\relax}%
\providecommand \BibitemShut  [1]{\csname bibitem#1\endcsname}%
\let\auto@bib@innerbib\@empty
\bibitem [{\citenamefont {Nisoli}, \citenamefont {Moessner},\ and\
  \citenamefont {Schiffer}(2013)}]{Nisoli2013}%
  \BibitemOpen
  \bibfield  {author} {\bibinfo {author} {\bibfnamefont {C.}~\bibnamefont
  {Nisoli}}, \bibinfo {author} {\bibfnamefont {R.}~\bibnamefont {Moessner}},\
  and\ \bibinfo {author} {\bibfnamefont {P.}~\bibnamefont {Schiffer}},\
  }\bibfield  {title} {\enquote {\bibinfo {title} {\textit{Colloquium} :
  Artificial spin ice: Designing and imaging magnetic frustration},}\ }\href
  {https://doi.org/10.1103/RevModPhys.85.1473} {\bibfield  {journal} {\bibinfo
  {journal} {Rev. Mod. Phys.}\ }\textbf {\bibinfo {volume} {85}},\ \bibinfo
  {pages} {1473--1490} (\bibinfo {year} {2013})}\BibitemShut {NoStop}%
\bibitem [{\citenamefont {Heyderman}\ and\ \citenamefont
  {Stamps}(2013)}]{Heyderman:2013gb}%
  \BibitemOpen
  \bibfield  {author} {\bibinfo {author} {\bibfnamefont {L.~J.}\ \bibnamefont
  {Heyderman}}\ and\ \bibinfo {author} {\bibfnamefont {R.~L.}\ \bibnamefont
  {Stamps}},\ }\bibfield  {title} {\enquote {\bibinfo {title} {{Artificial
  ferroic systems: novel functionality from structure, interactions and
  dynamics}},}\ }\href {https://doi.org/10.1088/0953-8984/25/36/363201}
  {\bibfield  {journal} {\bibinfo  {journal} {Journal of Physics: Condensed
  Matter}\ }\textbf {\bibinfo {volume} {25}},\ \bibinfo {pages} {363201}
  (\bibinfo {year} {2013})}\BibitemShut {NoStop}%
\bibitem [{\citenamefont {Nisoli}, \citenamefont {Kapaklis},\ and\
  \citenamefont {Schiffer}(2017)}]{Nisoli_NatPhys_Perspective}%
  \BibitemOpen
  \bibfield  {author} {\bibinfo {author} {\bibfnamefont {C.}~\bibnamefont
  {Nisoli}}, \bibinfo {author} {\bibfnamefont {V.}~\bibnamefont {Kapaklis}},\
  and\ \bibinfo {author} {\bibfnamefont {P.}~\bibnamefont {Schiffer}},\
  }\bibfield  {title} {\enquote {\bibinfo {title} {{Deliberate exotic magnetism
  via frustration and topology}},}\ }\href {https://doi.org/10.1038/nphys4059}
  {\bibfield  {journal} {\bibinfo  {journal} {Nature Physics}\ }\textbf
  {\bibinfo {volume} {13}},\ \bibinfo {pages} {200--203} (\bibinfo {year}
  {2017})}\BibitemShut {NoStop}%
\bibitem [{\citenamefont {Rougemaille}\ and\ \citenamefont
  {Canals}(2019)}]{Rougemaille_Colloquium_2019}%
  \BibitemOpen
  \bibfield  {author} {\bibinfo {author} {\bibfnamefont {N.}~\bibnamefont
  {Rougemaille}}\ and\ \bibinfo {author} {\bibfnamefont {B.}~\bibnamefont
  {Canals}},\ }\bibfield  {title} {\enquote {\bibinfo {title} {{Cooperative
  magnetic phenomena in artificial spin systems: spin liquids, Coulomb phase
  and fragmentation of magnetism {\textendash} a colloquium}},}\ }\href
  {https://doi.org/10.1140/epjb/e2018-90346-7} {\bibfield  {journal} {\bibinfo
  {journal} {Eur. Phys. J. B}\ }\textbf {\bibinfo {volume} {92}},\ \bibinfo
  {pages} {62} (\bibinfo {year} {2019})}\BibitemShut {NoStop}%
\bibitem [{\citenamefont {Arnalds}\ \emph {et~al.}(2016)\citenamefont
  {Arnalds}, \citenamefont {Chico}, \citenamefont {Stopfel}, \citenamefont
  {Kapaklis}, \citenamefont {B{\"a}renbold}, \citenamefont {Verschuuren},
  \citenamefont {Wolff}, \citenamefont {Neu}, \citenamefont {Bergman},\ and\
  \citenamefont {Hj{\"o}rvarsson}}]{Arnalds_2DIsing}%
  \BibitemOpen
  \bibfield  {author} {\bibinfo {author} {\bibfnamefont {U.~B.}\ \bibnamefont
  {Arnalds}}, \bibinfo {author} {\bibfnamefont {J.}~\bibnamefont {Chico}},
  \bibinfo {author} {\bibfnamefont {H.}~\bibnamefont {Stopfel}}, \bibinfo
  {author} {\bibfnamefont {V.}~\bibnamefont {Kapaklis}}, \bibinfo {author}
  {\bibfnamefont {O.}~\bibnamefont {B{\"a}renbold}}, \bibinfo {author}
  {\bibfnamefont {M.~A.}\ \bibnamefont {Verschuuren}}, \bibinfo {author}
  {\bibfnamefont {U.}~\bibnamefont {Wolff}}, \bibinfo {author} {\bibfnamefont
  {V.}~\bibnamefont {Neu}}, \bibinfo {author} {\bibfnamefont {A.}~\bibnamefont
  {Bergman}},\ and\ \bibinfo {author} {\bibfnamefont {B.}~\bibnamefont
  {Hj{\"o}rvarsson}},\ }\bibfield  {title} {\enquote {\bibinfo {title} {{A new
  look on the two-dimensional Ising model: thermal artificial spins}},}\ }\href
  {https://doi.org/10.1088/1367-2630/18/2/023008} {\bibfield  {journal}
  {\bibinfo  {journal} {New Journal of Physics}\ }\textbf {\bibinfo {volume}
  {18}},\ \bibinfo {pages} {023008} (\bibinfo {year} {2016})}\BibitemShut
  {NoStop}%
\bibitem [{\citenamefont {{\"O}stman}\ \emph
  {et~al.}(2018{\natexlab{a}})\citenamefont {{\"O}stman}, \citenamefont
  {Arnalds}, \citenamefont {Kapaklis}, \citenamefont {Taroni},\ and\
  \citenamefont {Hj{\"o}rvarsson}}]{Ostman_Ising_2018}%
  \BibitemOpen
  \bibfield  {author} {\bibinfo {author} {\bibfnamefont {E.}~\bibnamefont
  {{\"O}stman}}, \bibinfo {author} {\bibfnamefont {U.~B.}\ \bibnamefont
  {Arnalds}}, \bibinfo {author} {\bibfnamefont {V.}~\bibnamefont {Kapaklis}},
  \bibinfo {author} {\bibfnamefont {A.}~\bibnamefont {Taroni}},\ and\ \bibinfo
  {author} {\bibfnamefont {B.}~\bibnamefont {Hj{\"o}rvarsson}},\ }\bibfield
  {title} {\enquote {\bibinfo {title} {{Ising-like behaviour of mesoscopic
  magnetic chains}},}\ }\href {https://doi.org/10.1088/1361-648X/aad0c1}
  {\bibfield  {journal} {\bibinfo  {journal} {Journal of Physics: Condensed
  Matter}\ }\textbf {\bibinfo {volume} {30}},\ \bibinfo {pages} {365301}
  (\bibinfo {year} {2018}{\natexlab{a}})}\BibitemShut {NoStop}%
\bibitem [{\citenamefont {Wang}\ \emph {et~al.}(2006)\citenamefont {Wang},
  \citenamefont {Nisoli}, \citenamefont {Freitas}, \citenamefont {Li},
  \citenamefont {McConville}, \citenamefont {Cooley}, \citenamefont {Lund},
  \citenamefont {Samarth}, \citenamefont {Leighton}, \citenamefont {Crespi},\
  and\ \citenamefont {Schiffer}}]{Wang2006}%
  \BibitemOpen
  \bibfield  {author} {\bibinfo {author} {\bibfnamefont {R.~F.}\ \bibnamefont
  {Wang}}, \bibinfo {author} {\bibfnamefont {C.}~\bibnamefont {Nisoli}},
  \bibinfo {author} {\bibfnamefont {R.~S.}\ \bibnamefont {Freitas}}, \bibinfo
  {author} {\bibfnamefont {J.}~\bibnamefont {Li}}, \bibinfo {author}
  {\bibfnamefont {W.}~\bibnamefont {McConville}}, \bibinfo {author}
  {\bibfnamefont {B.~J.}\ \bibnamefont {Cooley}}, \bibinfo {author}
  {\bibfnamefont {M.~S.}\ \bibnamefont {Lund}}, \bibinfo {author}
  {\bibfnamefont {N.}~\bibnamefont {Samarth}}, \bibinfo {author} {\bibfnamefont
  {C.}~\bibnamefont {Leighton}}, \bibinfo {author} {\bibfnamefont {V.~H.}\
  \bibnamefont {Crespi}},\ and\ \bibinfo {author} {\bibfnamefont
  {P.}~\bibnamefont {Schiffer}},\ }\bibfield  {title} {\enquote {\bibinfo
  {title} {Artificial `spin ice' in a geometrically frustrated lattice of
  nanoscale ferromagnetic islands},}\ }\href
  {https://doi.org/10.1038/nature04447} {\bibfield  {journal} {\bibinfo
  {journal} {Nature (London)}\ }\textbf {\bibinfo {volume} {439}},\ \bibinfo
  {pages} {303--306} (\bibinfo {year} {2006})}\BibitemShut {NoStop}%
\bibitem [{\citenamefont {Perrin}, \citenamefont {Canals},\ and\ \citenamefont
  {Rougemaille}(2016)}]{Perrin_Nature_2016}%
  \BibitemOpen
  \bibfield  {author} {\bibinfo {author} {\bibfnamefont {Y.}~\bibnamefont
  {Perrin}}, \bibinfo {author} {\bibfnamefont {B.}~\bibnamefont {Canals}},\
  and\ \bibinfo {author} {\bibfnamefont {N.}~\bibnamefont {Rougemaille}},\
  }\bibfield  {title} {\enquote {\bibinfo {title} {{Extensive degeneracy,
  Coulomb phase and magnetic monopoles in artificial square ice}},}\ }\href
  {https://doi.org/10.1038/nature20155} {\bibfield  {journal} {\bibinfo
  {journal} {Nature}\ }\textbf {\bibinfo {volume} {540}},\ \bibinfo {pages}
  {410--413} (\bibinfo {year} {2016})}\BibitemShut {NoStop}%
\bibitem [{\citenamefont {{\"O}stman}\ \emph
  {et~al.}(2018{\natexlab{b}})\citenamefont {{\"O}stman}, \citenamefont
  {Stopfel}, \citenamefont {Chioar}, \citenamefont {Arnalds}, \citenamefont
  {Stein}, \citenamefont {Kapaklis},\ and\ \citenamefont
  {Hj{\"o}rvarsson}}]{Ostman_natphys_2018}%
  \BibitemOpen
  \bibfield  {author} {\bibinfo {author} {\bibfnamefont {E.}~\bibnamefont
  {{\"O}stman}}, \bibinfo {author} {\bibfnamefont {H.}~\bibnamefont {Stopfel}},
  \bibinfo {author} {\bibfnamefont {I.-A.}\ \bibnamefont {Chioar}}, \bibinfo
  {author} {\bibfnamefont {U.~B.}\ \bibnamefont {Arnalds}}, \bibinfo {author}
  {\bibfnamefont {A.}~\bibnamefont {Stein}}, \bibinfo {author} {\bibfnamefont
  {V.}~\bibnamefont {Kapaklis}},\ and\ \bibinfo {author} {\bibfnamefont
  {B.}~\bibnamefont {Hj{\"o}rvarsson}},\ }\bibfield  {title} {\enquote
  {\bibinfo {title} {{Interaction modifiers in artificial spin ices}},}\ }\href
  {https://doi.org/10.1038/s41567-017-0027-2} {\bibfield  {journal} {\bibinfo
  {journal} {Nature Physics}\ }\textbf {\bibinfo {volume} {14}},\ \bibinfo
  {pages} {375--379} (\bibinfo {year} {2018}{\natexlab{b}})}\BibitemShut
  {NoStop}%
\bibitem [{\citenamefont {Kapaklis}\ \emph {et~al.}(2014)\citenamefont
  {Kapaklis}, \citenamefont {Arnalds}, \citenamefont {Farhan}, \citenamefont
  {Chopdekar}, \citenamefont {Balan}, \citenamefont {Scholl}, \citenamefont
  {Heyderman},\ and\ \citenamefont {Hj\"orvarsson}}]{ThermalFluctuationsInASI}%
  \BibitemOpen
  \bibfield  {author} {\bibinfo {author} {\bibfnamefont {V.}~\bibnamefont
  {Kapaklis}}, \bibinfo {author} {\bibfnamefont {U.~B.}\ \bibnamefont
  {Arnalds}}, \bibinfo {author} {\bibfnamefont {A.}~\bibnamefont {Farhan}},
  \bibinfo {author} {\bibfnamefont {R.~V.}\ \bibnamefont {Chopdekar}}, \bibinfo
  {author} {\bibfnamefont {A.}~\bibnamefont {Balan}}, \bibinfo {author}
  {\bibfnamefont {A.}~\bibnamefont {Scholl}}, \bibinfo {author} {\bibfnamefont
  {L.~J.}\ \bibnamefont {Heyderman}},\ and\ \bibinfo {author} {\bibfnamefont
  {B.}~\bibnamefont {Hj\"orvarsson}},\ }\bibfield  {title} {\enquote {\bibinfo
  {title} {Thermal fluctuations in artificial spin ice},}\ }\href
  {https://doi.org/10.1038/nnano.2014.104} {\bibfield  {journal} {\bibinfo
  {journal} {Nature Nanotechnology}\ }\textbf {\bibinfo {volume} {9}},\
  \bibinfo {pages} {514--519} (\bibinfo {year} {2014})}\BibitemShut {NoStop}%
\bibitem [{\citenamefont {Andersson}\ \emph {et~al.}(2016)\citenamefont
  {Andersson}, \citenamefont {Pappas}, \citenamefont {Stopfel}, \citenamefont
  {\"Ostman}, \citenamefont {Stein}, \citenamefont {Nordblad}, \citenamefont
  {Mathieu}, \citenamefont {Hj\"orvarsson},\ and\ \citenamefont
  {Kapaklis}}]{Andersson2016}%
  \BibitemOpen
  \bibfield  {author} {\bibinfo {author} {\bibfnamefont {M.~S.}\ \bibnamefont
  {Andersson}}, \bibinfo {author} {\bibfnamefont {S.~D.}\ \bibnamefont
  {Pappas}}, \bibinfo {author} {\bibfnamefont {H.}~\bibnamefont {Stopfel}},
  \bibinfo {author} {\bibfnamefont {E.}~\bibnamefont {\"Ostman}}, \bibinfo
  {author} {\bibfnamefont {A.}~\bibnamefont {Stein}}, \bibinfo {author}
  {\bibfnamefont {P.}~\bibnamefont {Nordblad}}, \bibinfo {author}
  {\bibfnamefont {R.}~\bibnamefont {Mathieu}}, \bibinfo {author} {\bibfnamefont
  {B.}~\bibnamefont {Hj\"orvarsson}},\ and\ \bibinfo {author} {\bibfnamefont
  {V.}~\bibnamefont {Kapaklis}},\ }\bibfield  {title} {\enquote {\bibinfo
  {title} {Thermally induced magnetic relaxation in square artificial spin
  ice},}\ }\href {https://doi.org/10.1038/srep37097} {\bibfield  {journal}
  {\bibinfo  {journal} {Scientific Reports}\ }\textbf {\bibinfo {volume} {6}},\
  \bibinfo {pages} {37097} (\bibinfo {year} {2016})}\BibitemShut {NoStop}%
\bibitem [{\citenamefont {Morley}\ \emph {et~al.}(2017)\citenamefont {Morley},
  \citenamefont {Venero}, \citenamefont {Porro}, \citenamefont {Riley},
  \citenamefont {Stein}, \citenamefont {Steadman}, \citenamefont {Stamps},
  \citenamefont {Langridge},\ and\ \citenamefont {Marrows}}]{Morley2017}%
  \BibitemOpen
  \bibfield  {author} {\bibinfo {author} {\bibfnamefont {S.~A.}\ \bibnamefont
  {Morley}}, \bibinfo {author} {\bibfnamefont {D.~A.}\ \bibnamefont {Venero}},
  \bibinfo {author} {\bibfnamefont {J.~M.}\ \bibnamefont {Porro}}, \bibinfo
  {author} {\bibfnamefont {S.~T.}\ \bibnamefont {Riley}}, \bibinfo {author}
  {\bibfnamefont {A.}~\bibnamefont {Stein}}, \bibinfo {author} {\bibfnamefont
  {P.}~\bibnamefont {Steadman}}, \bibinfo {author} {\bibfnamefont {R.~L.}\
  \bibnamefont {Stamps}}, \bibinfo {author} {\bibfnamefont {S.}~\bibnamefont
  {Langridge}},\ and\ \bibinfo {author} {\bibfnamefont {C.~H.}\ \bibnamefont
  {Marrows}},\ }\bibfield  {title} {\enquote {\bibinfo {title}
  {Vogel-\relax{F}ulcher-\relax{T}ammann freezing of a thermally fluctuating
  artificial spin ice probed by x-ray photon correlation spectroscopy},}\
  }\href {https://doi.org/10.1103/physrevb.95.104422} {\bibfield  {journal}
  {\bibinfo  {journal} {Physical Review B}\ }\textbf {\bibinfo {volume} {95}},\
  \bibinfo {pages} {104422} (\bibinfo {year} {2017})}\BibitemShut {NoStop}%
\bibitem [{\citenamefont {Pohlit}\ \emph {et~al.}(2020)\citenamefont {Pohlit},
  \citenamefont {Muscas}, \citenamefont {Chioar}, \citenamefont {Stopfel},
  \citenamefont {Ciuciulkaite}, \citenamefont {\"Ostman}, \citenamefont
  {Pappas}, \citenamefont {Stein}, \citenamefont {Hj\"orvarsson}, \citenamefont
  {J\"onsson},\ and\ \citenamefont {Kapaklis}}]{ASI_susceptibility_2020}%
  \BibitemOpen
  \bibfield  {author} {\bibinfo {author} {\bibfnamefont {M.}~\bibnamefont
  {Pohlit}}, \bibinfo {author} {\bibfnamefont {G.}~\bibnamefont {Muscas}},
  \bibinfo {author} {\bibfnamefont {I.-A.}\ \bibnamefont {Chioar}}, \bibinfo
  {author} {\bibfnamefont {H.}~\bibnamefont {Stopfel}}, \bibinfo {author}
  {\bibfnamefont {A.}~\bibnamefont {Ciuciulkaite}}, \bibinfo {author}
  {\bibfnamefont {E.}~\bibnamefont {\"Ostman}}, \bibinfo {author}
  {\bibfnamefont {S.~D.}\ \bibnamefont {Pappas}}, \bibinfo {author}
  {\bibfnamefont {A.}~\bibnamefont {Stein}}, \bibinfo {author} {\bibfnamefont
  {B.}~\bibnamefont {Hj\"orvarsson}}, \bibinfo {author} {\bibfnamefont {P.~E.}\
  \bibnamefont {J\"onsson}},\ and\ \bibinfo {author} {\bibfnamefont
  {V.}~\bibnamefont {Kapaklis}},\ }\bibfield  {title} {\enquote {\bibinfo
  {title} {Collective magnetic dynamics in artificial spin ice probed by ac
  susceptibility},}\ }\href {https://doi.org/10.1103/PhysRevB.101.134404}
  {\bibfield  {journal} {\bibinfo  {journal} {Physical Review B}\ }\textbf
  {\bibinfo {volume} {101}},\ \bibinfo {pages} {134404} (\bibinfo {year}
  {2020})}\BibitemShut {NoStop}%
\bibitem [{\citenamefont {Bayer}\ \emph {et~al.}(2005)\citenamefont {Bayer},
  \citenamefont {Jorzick}, \citenamefont {Hillebrands}, \citenamefont
  {Demokritov}, \citenamefont {Kouba}, \citenamefont {Bozinoski}, \citenamefont
  {Slavin}, \citenamefont {Guslienko}, \citenamefont {Berkov}, \citenamefont
  {Gorn} \emph {et~al.}}]{bayer2005spin}%
  \BibitemOpen
  \bibfield  {author} {\bibinfo {author} {\bibfnamefont {C.}~\bibnamefont
  {Bayer}}, \bibinfo {author} {\bibfnamefont {J.}~\bibnamefont {Jorzick}},
  \bibinfo {author} {\bibfnamefont {B.}~\bibnamefont {Hillebrands}}, \bibinfo
  {author} {\bibfnamefont {S.}~\bibnamefont {Demokritov}}, \bibinfo {author}
  {\bibfnamefont {R.}~\bibnamefont {Kouba}}, \bibinfo {author} {\bibfnamefont
  {R.}~\bibnamefont {Bozinoski}}, \bibinfo {author} {\bibfnamefont
  {A.}~\bibnamefont {Slavin}}, \bibinfo {author} {\bibfnamefont {K.~Y.}\
  \bibnamefont {Guslienko}}, \bibinfo {author} {\bibfnamefont {D.}~\bibnamefont
  {Berkov}}, \bibinfo {author} {\bibfnamefont {N.}~\bibnamefont {Gorn}}, \emph
  {et~al.},\ }\bibfield  {title} {\enquote {\bibinfo {title} {Spin-wave
  excitations in finite rectangular elements of ni 80 fe 20},}\ }\href
  {https://doi.org/10.1103/PhysRevB.72.064427} {\bibfield  {journal} {\bibinfo
  {journal} {Physical Review B}\ }\textbf {\bibinfo {volume} {72}},\ \bibinfo
  {pages} {064427} (\bibinfo {year} {2005})}\BibitemShut {NoStop}%
\bibitem [{\citenamefont {Bessarab}, \citenamefont {Uzdin},\ and\ \citenamefont
  {J{\'{o}}nsson}(2012)}]{Bessarab2012}%
  \BibitemOpen
  \bibfield  {author} {\bibinfo {author} {\bibfnamefont {P.~F.}\ \bibnamefont
  {Bessarab}}, \bibinfo {author} {\bibfnamefont {V.~M.}\ \bibnamefont
  {Uzdin}},\ and\ \bibinfo {author} {\bibfnamefont {H.}~\bibnamefont
  {J{\'{o}}nsson}},\ }\bibfield  {title} {\enquote {\bibinfo {title} {Harmonic
  transition-state theory of thermal spin transitions},}\ }\href
  {https://doi.org/10.1103/physrevb.85.184409} {\bibfield  {journal} {\bibinfo
  {journal} {Physical Review B}\ }\textbf {\bibinfo {volume} {85}},\ \bibinfo
  {pages} {184409} (\bibinfo {year} {2012})}\BibitemShut {NoStop}%
\bibitem [{\citenamefont {Bessarab}, \citenamefont {Uzdin},\ and\ \citenamefont
  {J{\'{o}}nsson}(2013)}]{Bessarab2013}%
  \BibitemOpen
  \bibfield  {author} {\bibinfo {author} {\bibfnamefont {P.~F.}\ \bibnamefont
  {Bessarab}}, \bibinfo {author} {\bibfnamefont {V.~M.}\ \bibnamefont
  {Uzdin}},\ and\ \bibinfo {author} {\bibfnamefont {H.}~\bibnamefont
  {J{\'{o}}nsson}},\ }\bibfield  {title} {\enquote {\bibinfo {title} {Size and
  shape dependence of thermal spin transitions in nanoislands},}\ }\href
  {https://doi.org/10.1103/physrevlett.110.020604} {\bibfield  {journal}
  {\bibinfo  {journal} {Physical Review Letters}\ }\textbf {\bibinfo {volume}
  {110}},\ \bibinfo {pages} {020604} (\bibinfo {year} {2013})}\BibitemShut
  {NoStop}%
\bibitem [{\citenamefont {Gliga}\ \emph {et~al.}(2015)\citenamefont {Gliga},
  \citenamefont {K{\'a}kay}, \citenamefont {Heyderman}, \citenamefont
  {Hertel},\ and\ \citenamefont {Heinonen}}]{Gliga_PRB_2015}%
  \BibitemOpen
  \bibfield  {author} {\bibinfo {author} {\bibfnamefont {S.}~\bibnamefont
  {Gliga}}, \bibinfo {author} {\bibfnamefont {A.}~\bibnamefont {K{\'a}kay}},
  \bibinfo {author} {\bibfnamefont {L.~J.}\ \bibnamefont {Heyderman}}, \bibinfo
  {author} {\bibfnamefont {R.}~\bibnamefont {Hertel}},\ and\ \bibinfo {author}
  {\bibfnamefont {O.~G.}\ \bibnamefont {Heinonen}},\ }\bibfield  {title}
  {\enquote {\bibinfo {title} {{Broken vertex symmetry and finite zero-point
  entropy in the artificial square ice ground state}},}\ }\href
  {https://doi.org/10.1103/PhysRevB.92.060413} {\bibfield  {journal} {\bibinfo
  {journal} {Physical Review B}\ }\textbf {\bibinfo {volume} {92}},\ \bibinfo
  {pages} {060413} (\bibinfo {year} {2015})}\BibitemShut {NoStop}%
\bibitem [{\citenamefont {Gliga}\ \emph {et~al.}(2013)\citenamefont {Gliga},
  \citenamefont {K{\'a}kay}, \citenamefont {Hertel},\ and\ \citenamefont
  {Heinonen}}]{Gliga_PRL_2013}%
  \BibitemOpen
  \bibfield  {author} {\bibinfo {author} {\bibfnamefont {S.}~\bibnamefont
  {Gliga}}, \bibinfo {author} {\bibfnamefont {A.}~\bibnamefont {K{\'a}kay}},
  \bibinfo {author} {\bibfnamefont {R.}~\bibnamefont {Hertel}},\ and\ \bibinfo
  {author} {\bibfnamefont {O.~G.}\ \bibnamefont {Heinonen}},\ }\bibfield
  {title} {\enquote {\bibinfo {title} {{Spectral Analysis of Topological
  Defects in an Artificial Spin-Ice Lattice}},}\ }\href
  {https://doi.org/10.1103/PhysRevLett.110.117205} {\bibfield  {journal}
  {\bibinfo  {journal} {Physical Review Letters}\ }\textbf {\bibinfo {volume}
  {110}},\ \bibinfo {pages} {117205} (\bibinfo {year} {2013})}\BibitemShut
  {NoStop}%
\bibitem [{\citenamefont {Iacocca}\ \emph {et~al.}(2016)\citenamefont
  {Iacocca}, \citenamefont {Gliga}, \citenamefont {Stamps},\ and\ \citenamefont
  {Heinonen}}]{iacocca2016reconfigurable}%
  \BibitemOpen
  \bibfield  {author} {\bibinfo {author} {\bibfnamefont {E.}~\bibnamefont
  {Iacocca}}, \bibinfo {author} {\bibfnamefont {S.}~\bibnamefont {Gliga}},
  \bibinfo {author} {\bibfnamefont {R.~L.}\ \bibnamefont {Stamps}},\ and\
  \bibinfo {author} {\bibfnamefont {O.}~\bibnamefont {Heinonen}},\ }\bibfield
  {title} {\enquote {\bibinfo {title} {Reconfigurable wave band structure of an
  artificial square ice},}\ }\href {https://doi.org/10.1103/PhysRevB.93.134420}
  {\bibfield  {journal} {\bibinfo  {journal} {Physical Review B}\ }\textbf
  {\bibinfo {volume} {93}},\ \bibinfo {pages} {134420} (\bibinfo {year}
  {2016})}\BibitemShut {NoStop}%
\bibitem [{\citenamefont {Iacocca}, \citenamefont {Gliga},\ and\ \citenamefont
  {Heinonen}(2020)}]{iacocca2020tailoring}%
  \BibitemOpen
  \bibfield  {author} {\bibinfo {author} {\bibfnamefont {E.}~\bibnamefont
  {Iacocca}}, \bibinfo {author} {\bibfnamefont {S.}~\bibnamefont {Gliga}},\
  and\ \bibinfo {author} {\bibfnamefont {O.~G.}\ \bibnamefont {Heinonen}},\
  }\bibfield  {title} {\enquote {\bibinfo {title} {Tailoring spin-wave channels
  in a reconfigurable artificial spin ice},}\ }\href
  {https://doi.org/10.1103/PhysRevApplied.13.044047} {\bibfield  {journal}
  {\bibinfo  {journal} {Physical Review Applied}\ }\textbf {\bibinfo {volume}
  {13}},\ \bibinfo {pages} {044047} (\bibinfo {year} {2020})}\BibitemShut
  {NoStop}%
\bibitem [{\citenamefont {Vansteenkiste}\ \emph {et~al.}(2014)\citenamefont
  {Vansteenkiste}, \citenamefont {Leliaert}, \citenamefont {Dvornik},
  \citenamefont {Helsen}, \citenamefont {Garcia-Sanchez},\ and\ \citenamefont
  {Van~Waeyenberge}}]{mumax3}%
  \BibitemOpen
  \bibfield  {author} {\bibinfo {author} {\bibfnamefont {A.}~\bibnamefont
  {Vansteenkiste}}, \bibinfo {author} {\bibfnamefont {J.}~\bibnamefont
  {Leliaert}}, \bibinfo {author} {\bibfnamefont {M.}~\bibnamefont {Dvornik}},
  \bibinfo {author} {\bibfnamefont {M.}~\bibnamefont {Helsen}}, \bibinfo
  {author} {\bibfnamefont {F.}~\bibnamefont {Garcia-Sanchez}},\ and\ \bibinfo
  {author} {\bibfnamefont {B.}~\bibnamefont {Van~Waeyenberge}},\ }\bibfield
  {title} {\enquote {\bibinfo {title} {{The design and verification of
  MuMax3}},}\ }\href {https://doi.org/10.1063/1.4899186} {\bibfield  {journal}
  {\bibinfo  {journal} {AIP Advances}\ }\textbf {\bibinfo {volume} {4}},\
  \bibinfo {pages} {107133} (\bibinfo {year} {2014})}\BibitemShut {NoStop}%
\bibitem [{\citenamefont {Berkov}\ and\ \citenamefont
  {Gorn}(2005)}]{berkov2005stochastic}%
  \BibitemOpen
  \bibfield  {author} {\bibinfo {author} {\bibfnamefont {D.~V.}\ \bibnamefont
  {Berkov}}\ and\ \bibinfo {author} {\bibfnamefont {N.~L.}\ \bibnamefont
  {Gorn}},\ }\bibfield  {title} {\enquote {\bibinfo {title} {Stochastic dynamic
  simulations of fast remagnetization processes: recent advances and
  applications},}\ }\href {https://doi.org/10.1016/j.jmmm.2004.11.569}
  {\bibfield  {journal} {\bibinfo  {journal} {Journal of Magnetism and Magnetic
  Materials}\ }\textbf {\bibinfo {volume} {290}},\ \bibinfo {pages} {442--448}
  (\bibinfo {year} {2005})}\BibitemShut {NoStop}%
\bibitem [{\citenamefont {Brown~Jr}(1963)}]{brown1963thermal}%
  \BibitemOpen
  \bibfield  {author} {\bibinfo {author} {\bibfnamefont {W.~F.}\ \bibnamefont
  {Brown~Jr}},\ }\bibfield  {title} {\enquote {\bibinfo {title} {Thermal
  fluctuations of a single-domain particle},}\ }\href
  {https://doi.org/10.1103/PhysRev.130.1677} {\bibfield  {journal} {\bibinfo
  {journal} {Physical review}\ }\textbf {\bibinfo {volume} {130}},\ \bibinfo
  {pages} {1677} (\bibinfo {year} {1963})}\BibitemShut {NoStop}%
\bibitem [{\citenamefont {Leliaert}\ \emph {et~al.}(2017)\citenamefont
  {Leliaert}, \citenamefont {Mulkers}, \citenamefont {De~Clercq}, \citenamefont
  {Coene}, \citenamefont {Dvornik},\ and\ \citenamefont
  {Van~Waeyenberge}}]{Leliaert_thermal_mumax3}%
  \BibitemOpen
  \bibfield  {author} {\bibinfo {author} {\bibfnamefont {J.}~\bibnamefont
  {Leliaert}}, \bibinfo {author} {\bibfnamefont {J.}~\bibnamefont {Mulkers}},
  \bibinfo {author} {\bibfnamefont {J.}~\bibnamefont {De~Clercq}}, \bibinfo
  {author} {\bibfnamefont {A.}~\bibnamefont {Coene}}, \bibinfo {author}
  {\bibfnamefont {M.}~\bibnamefont {Dvornik}},\ and\ \bibinfo {author}
  {\bibfnamefont {B.}~\bibnamefont {Van~Waeyenberge}},\ }\bibfield  {title}
  {\enquote {\bibinfo {title} {{Adaptively time stepping the stochastic
  Landau-Lifshitz-Gilbert equation at nonzero temperature: Implementation and
  validation in MuMax3}},}\ }\href {https://doi.org/10.1063/1.5003957}
  {\bibfield  {journal} {\bibinfo  {journal} {AIP Advances}\ }\textbf {\bibinfo
  {volume} {7}},\ \bibinfo {pages} {125010} (\bibinfo {year}
  {2017})}\BibitemShut {NoStop}%
\bibitem [{\citenamefont {McMichael}\ and\ \citenamefont
  {Stiles}(2005)}]{mcmichael2005magnetic}%
  \BibitemOpen
  \bibfield  {author} {\bibinfo {author} {\bibfnamefont {R.~D.}\ \bibnamefont
  {McMichael}}\ and\ \bibinfo {author} {\bibfnamefont {M.~D.}\ \bibnamefont
  {Stiles}},\ }\bibfield  {title} {\enquote {\bibinfo {title} {Magnetic normal
  modes of nanoelements},}\ }\href {https://doi.org/10.1063/1.1852191}
  {\bibfield  {journal} {\bibinfo  {journal} {Journal of Applied Physics}\
  }\textbf {\bibinfo {volume} {97}},\ \bibinfo {pages} {10J901} (\bibinfo
  {year} {2005})}\BibitemShut {NoStop}%
\bibitem [{\citenamefont {Ciuciulkaite}\ \emph {et~al.}(2019)\citenamefont
  {Ciuciulkaite}, \citenamefont {{\"O}stman}, \citenamefont {Brucas},
  \citenamefont {Kumar}, \citenamefont {Verschuuren}, \citenamefont
  {Svedlindh}, \citenamefont {Hj{\"o}rvarsson},\ and\ \citenamefont
  {Kapaklis}}]{ciuciulkaite2019collective}%
  \BibitemOpen
  \bibfield  {author} {\bibinfo {author} {\bibfnamefont {A.}~\bibnamefont
  {Ciuciulkaite}}, \bibinfo {author} {\bibfnamefont {E.}~\bibnamefont
  {{\"O}stman}}, \bibinfo {author} {\bibfnamefont {R.}~\bibnamefont {Brucas}},
  \bibinfo {author} {\bibfnamefont {A.}~\bibnamefont {Kumar}}, \bibinfo
  {author} {\bibfnamefont {M.~A.}\ \bibnamefont {Verschuuren}}, \bibinfo
  {author} {\bibfnamefont {P.}~\bibnamefont {Svedlindh}}, \bibinfo {author}
  {\bibfnamefont {B.}~\bibnamefont {Hj{\"o}rvarsson}},\ and\ \bibinfo {author}
  {\bibfnamefont {V.}~\bibnamefont {Kapaklis}},\ }\bibfield  {title} {\enquote
  {\bibinfo {title} {Collective magnetization dynamics in nanoarrays of thin
  \relax{F}e\relax{P}d disks},}\ }\href
  {https://doi.org/10.1103/PhysRevB.99.184415} {\bibfield  {journal} {\bibinfo
  {journal} {Physical Review B}\ }\textbf {\bibinfo {volume} {99}},\ \bibinfo
  {pages} {184415} (\bibinfo {year} {2019})}\BibitemShut {NoStop}%
\bibitem [{\citenamefont {Sl{\"o}etjes}\ \emph {et~al.}(2019)\citenamefont
  {Sl{\"o}etjes}, \citenamefont {Digernes}, \citenamefont {Klewe},
  \citenamefont {Shafer}, \citenamefont {Li}, \citenamefont {Yang},
  \citenamefont {Qiu}, \citenamefont {Arenholz}, \citenamefont {Folven},
  \citenamefont {Grepstad} \emph {et~al.}}]{sloetjes2019effects}%
  \BibitemOpen
  \bibfield  {author} {\bibinfo {author} {\bibfnamefont {S.~D.}\ \bibnamefont
  {Sl{\"o}etjes}}, \bibinfo {author} {\bibfnamefont {E.}~\bibnamefont
  {Digernes}}, \bibinfo {author} {\bibfnamefont {C.}~\bibnamefont {Klewe}},
  \bibinfo {author} {\bibfnamefont {P.}~\bibnamefont {Shafer}}, \bibinfo
  {author} {\bibfnamefont {Q.}~\bibnamefont {Li}}, \bibinfo {author}
  {\bibfnamefont {M.}~\bibnamefont {Yang}}, \bibinfo {author} {\bibfnamefont
  {Z.}~\bibnamefont {Qiu}}, \bibinfo {author} {\bibfnamefont {E.}~\bibnamefont
  {Arenholz}}, \bibinfo {author} {\bibfnamefont {E.}~\bibnamefont {Folven}},
  \bibinfo {author} {\bibfnamefont {J.~K.}\ \bibnamefont {Grepstad}}, \emph
  {et~al.},\ }\bibfield  {title} {\enquote {\bibinfo {title} {Effects of
  lattice geometry on the dynamic properties of dipolar-coupled magnetic
  nanodisk arrays},}\ }\href {https://doi.org/10.1103/PhysRevB.99.064418}
  {\bibfield  {journal} {\bibinfo  {journal} {Physical Review B}\ }\textbf
  {\bibinfo {volume} {99}},\ \bibinfo {pages} {064418} (\bibinfo {year}
  {2019})}\BibitemShut {NoStop}%
\bibitem [{\citenamefont {Gurevich}\ and\ \citenamefont
  {Melkov}(1996)}]{gurevich1996magnetization}%
  \BibitemOpen
  \bibfield  {author} {\bibinfo {author} {\bibfnamefont {A.~G.}\ \bibnamefont
  {Gurevich}}\ and\ \bibinfo {author} {\bibfnamefont {G.~A.}\ \bibnamefont
  {Melkov}},\ }\href@noop {} {\emph {\bibinfo {title} {Magnetization
  oscillations and waves}}}\ (\bibinfo  {publisher} {CRC press},\ \bibinfo
  {year} {1996})\BibitemShut {NoStop}%
\bibitem [{\citenamefont {R{\"u}ckriegel}\ and\ \citenamefont
  {Kopietz}(2015)}]{ruckriegel2015rayleigh}%
  \BibitemOpen
  \bibfield  {author} {\bibinfo {author} {\bibfnamefont {A.}~\bibnamefont
  {R{\"u}ckriegel}}\ and\ \bibinfo {author} {\bibfnamefont {P.}~\bibnamefont
  {Kopietz}},\ }\bibfield  {title} {\enquote {\bibinfo {title}
  {Rayleigh-\relax{J}eans condensation of pumped magnons in thin-film
  ferromagnets},}\ }\href {https://doi.org/10.1103/PhysRevLett.115.157203}
  {\bibfield  {journal} {\bibinfo  {journal} {Physical Review Letters}\
  }\textbf {\bibinfo {volume} {115}},\ \bibinfo {pages} {157203} (\bibinfo
  {year} {2015})}\BibitemShut {NoStop}%
\bibitem [{\citenamefont {Ding}\ \emph {et~al.}(2005)\citenamefont {Ding},
  \citenamefont {Schmid}, \citenamefont {Li}, \citenamefont {Guslienko},\ and\
  \citenamefont {Bader}}]{ding_magnetic_2005}%
  \BibitemOpen
  \bibfield  {author} {\bibinfo {author} {\bibfnamefont {H.~F.}\ \bibnamefont
  {Ding}}, \bibinfo {author} {\bibfnamefont {A.~K.}\ \bibnamefont {Schmid}},
  \bibinfo {author} {\bibfnamefont {D.}~\bibnamefont {Li}}, \bibinfo {author}
  {\bibfnamefont {K.~Y.}\ \bibnamefont {Guslienko}},\ and\ \bibinfo {author}
  {\bibfnamefont {S.~D.}\ \bibnamefont {Bader}},\ }\bibfield  {title} {\enquote
  {\bibinfo {title} {Magnetic {Bistability} of {Co} {Nanodots}},}\ }\href
  {https://doi.org/10.1103/PhysRevLett.94.157202} {\bibfield  {journal}
  {\bibinfo  {journal} {Physical Review Letters}\ }\textbf {\bibinfo {volume}
  {94}},\ \bibinfo {pages} {157202} (\bibinfo {year} {2005})}\BibitemShut
  {NoStop}%
\bibitem [{\citenamefont {Semenova}, \citenamefont {Berkov},\ and\
  \citenamefont {Gorn}(2020)}]{Semenova_PRB_2020}%
  \BibitemOpen
  \bibfield  {author} {\bibinfo {author} {\bibfnamefont {E.~K.}\ \bibnamefont
  {Semenova}}, \bibinfo {author} {\bibfnamefont {D.~V.}\ \bibnamefont
  {Berkov}},\ and\ \bibinfo {author} {\bibfnamefont {N.~L.}\ \bibnamefont
  {Gorn}},\ }\bibfield  {title} {\enquote {\bibinfo {title} {{Evaluation of the
  switching rate for magnetic nanoparticles: Analysis, optimization, and
  comparison of various numerical simulation algorithms}},}\ }\href
  {https://doi.org/10.1103/PhysRevB.102.144419} {\bibfield  {journal} {\bibinfo
   {journal} {Physical Review B}\ }\textbf {\bibinfo {volume} {102}},\ \bibinfo
  {pages} {144419} (\bibinfo {year} {2020})}\BibitemShut {NoStop}%
\bibitem [{\citenamefont {Mellado}(2020)}]{Mellado_arXiv_2020}%
  \BibitemOpen
  \bibfield  {author} {\bibinfo {author} {\bibfnamefont {P.}~\bibnamefont
  {Mellado}},\ }\bibfield  {title} {\enquote {\bibinfo {title} {Timescales in
  the thermal dynamics of magnetic dipolar clusters},}\ }\href
  {https://doi.org/10.1103/PhysRevB.102.214442} {\bibfield  {journal} {\bibinfo
   {journal} {Phys. Rev. B}\ }\textbf {\bibinfo {volume} {102}},\ \bibinfo
  {pages} {214442} (\bibinfo {year} {2020})}\BibitemShut {NoStop}%
\bibitem [{\citenamefont {Leo}\ \emph {et~al.}(2021)\citenamefont {Leo},
  \citenamefont {Pancaldi}, \citenamefont {Koraltan}, \citenamefont
  {Gonz{\'a}lez}, \citenamefont {Abert}, \citenamefont {Vogler}, \citenamefont
  {Slanovc}, \citenamefont {Bruckner}, \citenamefont {Heistracher},
  \citenamefont {Hofhuis} \emph {et~al.}}]{leo2021chiral}%
  \BibitemOpen
  \bibfield  {author} {\bibinfo {author} {\bibfnamefont {N.}~\bibnamefont
  {Leo}}, \bibinfo {author} {\bibfnamefont {M.}~\bibnamefont {Pancaldi}},
  \bibinfo {author} {\bibfnamefont {S.}~\bibnamefont {Koraltan}}, \bibinfo
  {author} {\bibfnamefont {P.~V.}\ \bibnamefont {Gonz{\'a}lez}}, \bibinfo
  {author} {\bibfnamefont {C.}~\bibnamefont {Abert}}, \bibinfo {author}
  {\bibfnamefont {C.}~\bibnamefont {Vogler}}, \bibinfo {author} {\bibfnamefont
  {F.}~\bibnamefont {Slanovc}}, \bibinfo {author} {\bibfnamefont
  {F.}~\bibnamefont {Bruckner}}, \bibinfo {author} {\bibfnamefont
  {P.}~\bibnamefont {Heistracher}}, \bibinfo {author} {\bibfnamefont
  {K.}~\bibnamefont {Hofhuis}}, \emph {et~al.},\ }\bibfield  {title} {\enquote
  {\bibinfo {title} {Chiral switching and dynamic barrier reductions in
  artificial square ice},}\ }\href@noop {} {\bibfield  {journal} {\bibinfo
  {journal} {New Journal of Physics}\ } (\bibinfo {year} {2021})}\BibitemShut
  {NoStop}%
\bibitem [{\citenamefont {Hofhuis}\ \emph {et~al.}(2020)\citenamefont
  {Hofhuis}, \citenamefont {Hrabec}, \citenamefont {Arava}, \citenamefont
  {Leo}, \citenamefont {Huang}, \citenamefont {Chopdekar}, \citenamefont
  {Parchenko}, \citenamefont {Kleibert}, \citenamefont {Koraltan},
  \citenamefont {Abert}, \citenamefont {Vogler}, \citenamefont {Suess},
  \citenamefont {Derlet},\ and\ \citenamefont
  {Heyderman}}]{Hofhuis_DM_ASI_2020}%
  \BibitemOpen
  \bibfield  {author} {\bibinfo {author} {\bibfnamefont {K.}~\bibnamefont
  {Hofhuis}}, \bibinfo {author} {\bibfnamefont {A.}~\bibnamefont {Hrabec}},
  \bibinfo {author} {\bibfnamefont {H.}~\bibnamefont {Arava}}, \bibinfo
  {author} {\bibfnamefont {N.}~\bibnamefont {Leo}}, \bibinfo {author}
  {\bibfnamefont {Y.-L.}\ \bibnamefont {Huang}}, \bibinfo {author}
  {\bibfnamefont {R.~V.}\ \bibnamefont {Chopdekar}}, \bibinfo {author}
  {\bibfnamefont {S.}~\bibnamefont {Parchenko}}, \bibinfo {author}
  {\bibfnamefont {A.}~\bibnamefont {Kleibert}}, \bibinfo {author}
  {\bibfnamefont {S.}~\bibnamefont {Koraltan}}, \bibinfo {author}
  {\bibfnamefont {C.}~\bibnamefont {Abert}}, \bibinfo {author} {\bibfnamefont
  {C.}~\bibnamefont {Vogler}}, \bibinfo {author} {\bibfnamefont
  {D.}~\bibnamefont {Suess}}, \bibinfo {author} {\bibfnamefont {P.~M.}\
  \bibnamefont {Derlet}},\ and\ \bibinfo {author} {\bibfnamefont {L.~J.}\
  \bibnamefont {Heyderman}},\ }\bibfield  {title} {\enquote {\bibinfo {title}
  {{Thermally superactive artificial kagome spin ice structures obtained with
  the interfacial Dzyaloshinskii-Moriya interaction}},}\ }\href
  {https://doi.org/10.1103/PhysRevB.102.180405} {\bibfield  {journal} {\bibinfo
   {journal} {Physical Review B}\ }\textbf {\bibinfo {volume} {102}},\ \bibinfo
  {pages} {180405} (\bibinfo {year} {2020})}\BibitemShut {NoStop}%
\bibitem [{\citenamefont {Lendinez}\ and\ \citenamefont
  {Jungfleisch}(2020)}]{Lendinez_review_2020}%
  \BibitemOpen
  \bibfield  {author} {\bibinfo {author} {\bibfnamefont {S.}~\bibnamefont
  {Lendinez}}\ and\ \bibinfo {author} {\bibfnamefont {M.~B.}\ \bibnamefont
  {Jungfleisch}},\ }\bibfield  {title} {\enquote {\bibinfo {title}
  {{Magnetization dynamics in artificial spin ice}},}\ }\href
  {https://doi.org/10.1088/1361-648X/ab3e78} {\bibfield  {journal} {\bibinfo
  {journal} {Journal of Physics: Condensed Matter}\ }\textbf {\bibinfo {volume}
  {32}},\ \bibinfo {pages} {013001} (\bibinfo {year} {2020})}\BibitemShut
  {NoStop}%
\bibitem [{\citenamefont {Gypens}, \citenamefont {Leliaert},\ and\
  \citenamefont {Van~Waeyenberge}(2018)}]{Gypens_2018}%
  \BibitemOpen
  \bibfield  {author} {\bibinfo {author} {\bibfnamefont {P.}~\bibnamefont
  {Gypens}}, \bibinfo {author} {\bibfnamefont {J.}~\bibnamefont {Leliaert}},\
  and\ \bibinfo {author} {\bibfnamefont {B.}~\bibnamefont {Van~Waeyenberge}},\
  }\bibfield  {title} {\enquote {\bibinfo {title} {{Balanced Magnetic Logic
  Gates in a Kagome Spin Ice}},}\ }\href
  {https://doi.org/10.1103/PhysRevApplied.9.034004} {\bibfield  {journal}
  {\bibinfo  {journal} {Physical Review Applied}\ }\textbf {\bibinfo {volume}
  {9}},\ \bibinfo {pages} {034004} (\bibinfo {year} {2018})}\BibitemShut
  {NoStop}%
\end{thebibliography}
\end{document}